\documentclass[amssymb,amsmath,pra,twocolumn,floatfix,showpacs,10pt]{revtex4}
\usepackage{graphicx}

\begin{document}
\title{Intensity-dependent dispersion under conditions of electromagnetically
induced transparency in coherently prepared multi-state atoms.}
\author{Andrew D. Greentree$^{1, 2}$}
\author{Derek Richards$^{3}$}
\author{J. A. Vaccaro$^{4}$}
\author{A. V. Durrant$^{1}$}
\author{S. R. de Echaniz$^{5}$}
\author{D. M. Segal$^{5}$}
\author{J. P. Marangos$^{5}$}
\affiliation{$^{1}$Quantum Processes Group, Department of Physics
and Astronomy, The Open University, Walton Hall, Milton-Keynes,
MK7 6AA, United Kingdom}

\affiliation{$^2$ Centre for Quantum Computer Technology, School
of Physics, The University of New South Wales, Sydney, NSW 2052,
Australia}

\affiliation {$^{3}$Quantum Processes Group, Department of Applied
Mathematics, The Open University, Walton Hall, Milton-Keynes, MK7
6AA, United Kingdom}

\affiliation{$^{4}$Department of Physics and Astronomy, The
University of Hertfordshire, College Lane, Hatfield AL10 9AB,
United Kingdom}

\affiliation{$^{5}$Quantum Optics and Laser Science, Blackett
Laboratory, Imperial College of Science, Technology and Medicine,
Prince Consort Road, London SW7 2BW, United Kingdom}

\date{\today}

\begin{abstract}
Interest in lossless nonlinearities has focussed on the the
dispersive properties of $\Lambda $ systems under conditions of
electromagnetically induced transparency (EIT). \ We generalize
the lambda system by introducing further degenerate states to
realize a `Chain $ \Lambda $' atom where multiple coupling of the
probe field significantly enhances the intensity dependent
dispersion without compromising the EIT condition.
\end{abstract}


\maketitle

There has been much interest given lately to the enhancement of
optical nonlinearities in Electromagnetically Induced Transparency
(EIT). \ Most of the work has focussed on the three-state system
in the $\Lambda $ configuration, which has provided some dramatic
examples of nonlinear optical effects. \ Examples include
ultra-slow \cite{bib:UltraSlow}, stopped \cite{bib:StoppedLight}
and superluminal \cite{bib:UltraFast} group velocities, coherent
sideband generation \cite{bib:CoherentSideband} etc. All these
nonlinear processes depend on the creation of coherent
superpositions of the ground states with accompanying loss of
absorption, and such mechanisms were described in
\cite{bib:Nonlinear}. \ Thorough reviews of EIT and its properties
can be found in \cite{bib:Marangos1998} and
\cite{bib:MatskoReview2001}.

Recent investigations of nonlinear optics at the few or single
photon levels have identified four state systems where the probe
field simultaneously couples two transitions in the {\sf N}
configuration. \ Examples of applications for such work include
photon blockade \cite {bib:PhotonBlockade} and two photon
absorptive switches \cite {bib:AbsorptiveQSwitch}. \ The classical
precursors to such experiments have also been performed
\cite{bib:ATVee}\cite{bib:deEchaniz2001}. \ Other experiments on
the {\sf N} scheme have been performed by \'{E}ntin {\it et al.
}\cite{bib:Entin2000}. \ In order to realize larger nonlinear
effects, Zubairy {\it et al}. \cite{bib:Zubairy2002} suggested an
extension where the more usual {\sf N} configuration was extended
to a system with an arbitrary (even) number of states where all
the states are resonantly coupled except on the final transition
where detuning is present.  \ This scheme shows enhanced
nonlinearities of not only $\chi^3$ but also higher order
susceptibilities. \ One problem with this scheme and the standard
{\sf N} scheme is to do with the need to balance the required
nonlinearity and decoherence in the system. \ To enhance the
nonlinearity present in the systems it is important for the
detuning of the final probe field to be minimized, however
decreasing the detuning increases the amount of the final excited
state which is mixed into the coherent superposition state,
resulting in an increase in decoherence and optical losses. \ This
problem is to some extent circumvented by the absorptive switch of
Harris and Yamomoto \cite{bib:AbsorptiveQSwitch} by exploiting
such losses, and in photon blockade by using the cavity
nonlinearity to prevent absorption of the final photon. \ Still
the increase in decoherence proves to be a difficulty in
experimental precursors to these processes and causes problems in
travelling wave configurations.

An alternative multistate configuration for investigating EIT
enhanced nonlinearities is the tripod configuration, studied
recently by Paspalakis and Knight \cite{bib:Paspalakis2002} and
earlier considered by Morris and Shore \cite{bib:Morris1983}. This
system has many of the advantages of the {\sf N} system, but by
maintaining superposition states of the three ground states also
avoids the excess decoherence of the {\sf N} system.  Morris and
Shore \cite{bib:Morris1983} also mentioned a multi-leg extension
of the tripod scheme.

Here we present an alternative extension to the standard $\Lambda$
configuration which we term as the Chain $\Lambda$ configuration,
depicted in Fig. 1. \ We start with a $\Lambda $ atom [Fig. 1(a)]
with ground states $|g_1\rangle$ and $|g_2\rangle$ and excited
state $|e_1\rangle$.  The $g_{1}-e_{1}$ $\left( g_{2}-e_{1}\right)
$transition is excited by a probe (coupling) field of frequency
$\omega _{p}$ $\left( \omega _{c}\right) $ and detuning $\Delta_p
= \omega _p-\omega _{e_{1}g_{1}}$ $\left( \Delta _{c}=\omega
_{c}-\omega _{e_{1}g_{2}}\right) $. \ The probe (coupling) Rabi
frequency is $P$ $\left( C\right) $.\ \ We assume for convenience
that the transition frequencies $\omega _{e_{1}g_{1}}$ and $\omega
_{e_{1}g_{2}}$ are equal and that appropriate selection rules
ensure that the probe and coupling fields only interact with their
designated states (a concrete example of how this can be achieved
is described below).

\begin{figure}[tb!]
\includegraphics[width=\columnwidth,clip]{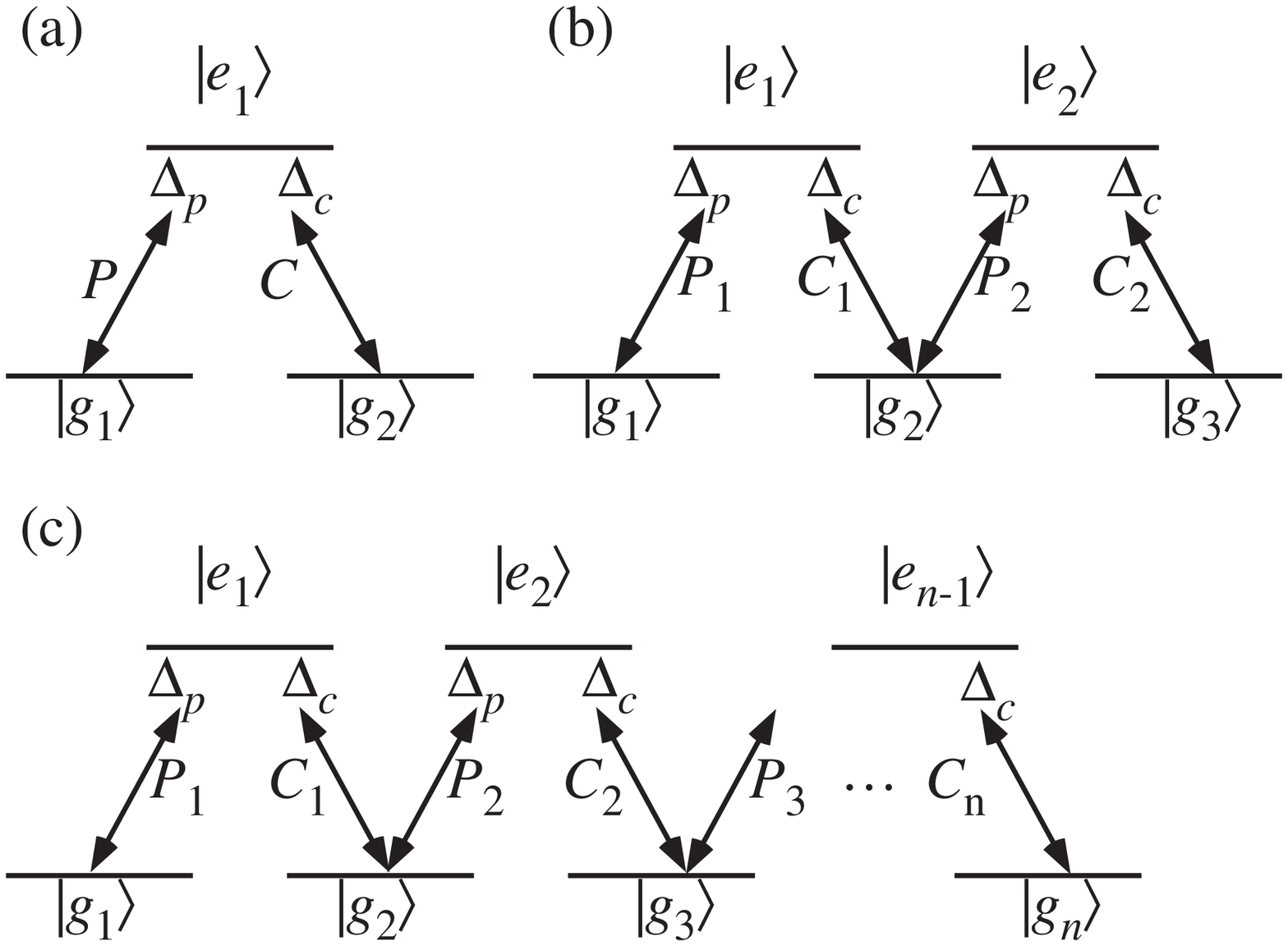}
\caption{\label{fig1} Energy level configurations for Chain
$\Lambda $ atoms. \ (a) is the usual $\Lambda $ system, (b) is the
5 state Chain $\Lambda $ or {\sf M} system, (c) shows the
generalization to higher number of states.}
\end{figure}

The five-state Chain $\Lambda$ is an {\sf M} system and is
illustrated in Fig. 1(b). \ Here the ground states are labelled
$|g_1\rangle$, $|g_2\rangle$, $|g_3\rangle$, and the excited
states $|e_1\rangle$, $|e_{2}\rangle$. \ The probe field excites
the $g_{1}-e_{1}$ and $g_{2}-e_{2}$ transitions simultaneously
with Rabi frequencies $P_{1}$ and $P_{2}$ respectively (different
labels are applied to take account of the different coupling
strengths of the transitions), whilst the coupling field excites
the $g_{2}-e_{1}$ and $g_{3}-e_{2}$ transitions with Rabi
frequencies $C_{1}$ and $C_{2}$. \ We note that a complementary
work by Matsko {\it et al.} \cite{bib:MatskoPreprint2002}, which
looks at Faraday rotation and Kerr nonlinearities in the
$\Lambda$, {\sf N} and {\sf M} schemes, has been performed which
confirms some of our predictions about nonlinearities in these
systems.  An early study of the {\sf M} scheme in the context of
degenerate two level systems was also performed by Morris and
Shore \cite{bib:Morris1983}

The Chain Lambda atom with $2n-1$ states is shown in Fig. 1(c). \
The ground states are denoted $|g_1\rangle$, ..., $|g_n\rangle$,
and the excited states $|e_1\rangle$, ..., $|e_{n-1}\rangle$. \
The probe field excites the $g_{j}-e_{j}$ transition with Rabi
frequency $P_{j}$ and the coupling field excites the
$g_{j+1}-e_{j}$ transition with Rabi frequency $C_{j}$.

In order to gain insight into the problem we consider first the
Hamiltonian for the Chain $\Lambda $ atom with $2n-1$ states. \
Using the rotating wave approximation this can be written
\begin{eqnarray*}
\frac{{\cal H}}{\hbar } &=&
    \sum_{j=1}^{n}\left( j-1\right) \left(
\Delta _{p}-\Delta _{c}\right) \left| g_{j}\right\rangle
\left\langle g_{j}\right| \\
&&+ \sum_{j=1}^{n-1}\left[ j\Delta _{p}-\left( j-1\right) \Delta
_{c}\right]
\left| e_{j}\right\rangle \left\langle e_{j}\right|  \\
&&+\sum_{j=1}^{n-1}P_{j}\left( \left| g_{j}\right\rangle
\left\langle e_{j}\right| +\left| e_{j}\right\rangle \left\langle
g_{j}\right| \right)  \\
&&+\sum_{j=1}^{n-1}C_{j}\left( \left| g_{j+1}\right\rangle
\left\langle e_{j}\right| +\left| e_{j}\right\rangle \left\langle
g_{j+1}\right| \right) .
\end{eqnarray*}
We ignore the decay from the excited states in our analytical
analysis in order to gain simple expressions for the dressed
states of the field-atoms system and thus gain a clearer
understanding of the problem. \ Furthermore we shall concentrate
our analysis on the optical nonlinearities which are present in
the vicinity of the dark state, which are relatively insensitive
to decay.  This is evident by direct comparisons between numerical
solutions of the complete master equation with decay, and our
decay-free analytic expressions. \ The Hamiltonian can be
conveniently expressed as a tridiagonal matrix with state ordering
$\left| g_{1}\right\rangle $, $\left| e_{1}\right\rangle $, $
\left| g_{2}\right\rangle $, ..., $\left| g_{n}\right\rangle $,
\[
\frac{{\cal H}}{\hbar }=\left[
\begin{array}{ccccccc}
0 & P_{1} & 0 & 0 &  &  &  \\
P_{1} & \Delta _{p} & C_{1} & 0 &  &  &  \\
0 & C_{1} & \Delta_{cp} & P_{2} &  &  &  \\
0 & 0 & P_{2} & \ddots  & \ddots  &  &  \\
&  &  & \ddots  & \left( n-1\right) \Delta_{cp}  &
P_{n} & 0 \\
&  &  &  & P_{n} & n \Delta_{cp} + \Delta _{c} & C_{n} \\
&  &  &  & 0 & C_{n} & n\Delta_{cp}
\end{array}
\right],
\]
where we have introduced $\Delta_{cp} = \Delta_p - \Delta_c$.

Following the approach taken by Kuang {\it et al.}
\cite{bib:Kuangpreprint} and Zubairy {\it et al.}
\cite{bib:Zubairy2002}, we first calculate the eigenvectors of the
Hamiltonian. \ These can be written as
\[
\left| {\cal D}_{i}\right\rangle =\alpha _{i,g_1}\left|
g_{1}\right\rangle +\alpha _{i,e_1}\left| e_{1}\right\rangle
+\alpha _{i,g_2}\left| g_{2}\right\rangle +\cdots +\alpha
_{i,g_n}\left| g_{n}\right\rangle ,
\]
where $i$ varies from $1$ to $(2n-1)$. \ It is clearly not
possible to give general solutions for the $\alpha _{i}$'s for
atoms with more than 3 states, although one may simply derive
numerical results. \ However, if we invoke the adiabatic
hypothesis \cite {bib:Kuangpreprint} and assume that the probe
detuning is small $\left( \Delta _{p} \ll P_{i},C_{i}\right) $,
the probe field is turned on slowly, and the coupling field
resonant, then we may assume that the system evolves {\em solely}
into the dressed state with energy closest to 0. \ For convenience
we denote this state $\left| {\cal D}_{0}^{(n)}\right\rangle $
where $n$ is the number of states in the Chain $\Lambda$ atom. \
Using MAPLE \cite{bib:Maple}, and the simplification that
$P_{i}=P,$ $C_{i}=C$, $i=1,2\ldots n$ we have derived expressions
for $\left| {\cal D}_{0}^{(n)}\right\rangle $. \ These are
presented in table \ref{tab:Table1} as unnormalized quantities and
where
$\Omega ^{2}=C^{2}+P^{2}$. \ Note that the results for $\left| {\cal D}%
_{0}^{\left( 3\right) }\right\rangle $ are equivalent to those
which appear in Ref. \cite{bib:Kuangpreprint} and were also used
in Ref. \cite{bib:deEchaniz2001}.

\begin{table}
\begin{tabular}{||c||c|c|c|}
\hline\hline
& $\left| {\cal D}_{0}^{\left( 3\right) }\right\rangle $ & $\left| {\cal D}%
_{0}^{\left( 5\right) }\right\rangle $ & $\left| {\cal
D}_{0}^{\left( 7\right) }\right\rangle $ \\ \hline\hline
$\alpha _{0,g_1}$ & $\frac{C}{P}$ & $\frac{C^{2}}{P^{2}}$ & $-\frac{C^{3}}{%
P^{3}}$ \\
$\alpha _{0,e_1}$ & $\Delta_{p}\frac{C}{\Omega ^{2}}$ &
$\Delta_{p}\frac{C^2 (\Omega^2+P^2)} {P(\Omega^4-C^2P^2)}$ &
$\Delta_{p}\frac{C(\Omega^2-P^2)(\Omega^2+2P^2)}{\Omega^2(\Omega^4-C^2P^2)}$ \\
$\alpha _{0,g_2}$ & $-1$ & $-\frac{C}{P}$ & $\frac{C^{2}}{P^{2}}$ \\
$\alpha _{0,e_2}$ & $-$ &
$\Delta_{p}\frac{C(2\Omega^2-P^2)}{\Omega ^{4}-C^2P^2}$ &
$\Delta_{p}\frac{2C^{2}\Omega^{2}}{P\left( \Omega ^{4}-C^{2}P^{2}\right) }$ \\
$\alpha _{0,g_3}$ & $-$ & $1$ & $-\frac{C}{P}$ \\
$\alpha _{0,e_3}$ & $-$ & $-$ & $\Delta _{p}\frac{C\left( \Omega
^{4}+2C^{4}\right) }{\Omega ^{2}\left( 2C^{2}P^{2}-\Omega ^{4}\right) }$ \\
$\alpha _{0,g_4}$ & $-$ & $-$ & $1$ \\ \hline
\end{tabular}
\caption{\label{tab:Table1} Unnormalized coefficients of $|{\cal
D}_0\rangle $ for 3, 5 and 7 state Chain $\Lambda $ atoms.}
\end{table}

Short of directly creating an artificial atomic structure in a
quantum well type material, or in an optical lattice, it is
important to investigate whether the required Chain $ \Lambda $
structure is naturally present in any materials. \ A simple
approximate realization to the Chain $ \Lambda $ configuration is
obtained by exciting an $F=n$ to $F^{\prime }=n$ transition in an
atomic vapor where, as before, $2n-1$ is the number of states in
the Chain $ \Lambda $. \ We illustrate this for the five-state
Chain $\Lambda $ atom in Fig. 2, although the scheme generalizes
in an obvious manner. \ We assume that the coupling field is
$\sigma^-$ polarized (and therefore excites transitions from
$m_{z}=n$ to $m_{z}=n-1$) and the probe $\sigma^+$ polarized
(exciting transitions from $m_{z}=n$ to $m_{z}=n+1$). \ Notice
that without any preparation there are two systems here, first the
required {\sf M} system (bold lines in Fig. 2), and second an
undesired {\sf W} system (dashed lines). \ In order to select the
{\sf M} over the {\sf W} we first apply the coupling field, this
has the effect of optically pumping the population into the
$m_{z}=-2$ state. Next the probe beam is turned on sufficiently
slowly to ensure that the system evolves adiabatically to the
desired dark state.  We also note that the {\sf W} system does not
have any dark states.  So even without the adiabatic state
preparation, any population in the {\sf W} system will eventually
be optically pumped into the desired dark state of the {\sf M}
system.  We also note that the presence of {\sf M} systems have
been identified in conjunction with $\Lambda$ systems at least
twice before in Refs. \cite{bib:Morris1983}
\cite{bib:MatskoPreprint2002}. It is important to notice that our
method of realizing Chain $\Lambda$ systems and performing
experiments with them, is not significantly more complex than
standard experiments on simple $\Lambda$ systems, all that is
necessary to achieve the enhanced nonlinearities is the
appropriate choice of transition.

\begin{figure}[tb!]
\includegraphics[width=\columnwidth,clip]{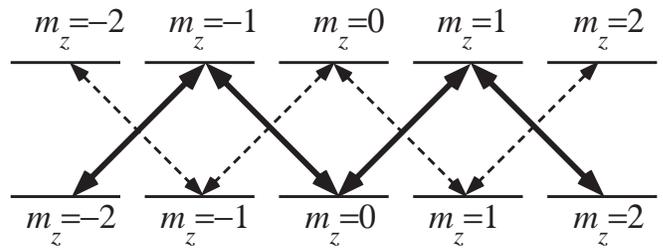}
\caption{\label{fig2} A possible way to realize the {\sf M} system
using an $F=2$ to $F^{\prime}=2$ transition. \ The coupling field
is $\sigma^-$ polarized and turned on before the $\sigma^+$
polarized probe. \ In this way the desired {\sf M} system (bold)
is favored over the {\sf W} system (dashed).}
\end{figure}

To study intensity-dependent dispersion it is necessary to extract
the susceptibility at the probe frequency as a function of small
probe detuning
\[
\chi =\kappa \sum_{j=1}^{n}\frac{\mu _{gjej}^{2}}{P_{j}}\rho
_{g_{j}e_{j}}=\kappa \sum_{j=1}^{n}\frac{\mu
_{gjej}^{2}}{P_{j}}\alpha _{0,g_j}^{\ast }\alpha _{0,e_j},
\]
where $\kappa =2\pi {\cal N}/\epsilon _{0}\hbar $, ${\cal N}$
being the atomic density and $\mu _{g_{j}e_{j}}$ the dipole moment
of the $g_{j}-e_{j}$ transition. The $\ast $ denotes complex
conjugation. \ In order to calculate realistic values of the
dispersion which would be attainable in standard experiments with
alkali atoms (for example in a magneto-optical trap or vapor
cell), we have taken ${\cal N}=3\times 10^{15} {\rm m}^{-3}$, $\mu
_{g_{j}e_{j}}=2\times 10^{-29} {\rm Cm}$ and $\Gamma = 5.6 {\rm
MHz}$.

In order to derive simple results for the nonlinear dispersions,
we shall assume the coupling constants for all transitions to be
equal, i.e. $\mu _{g_{i}e_{i}}=\mu $ and so the probe (coupling)
field Rabi frequencies are the same for all transitions $P_{i}=P$
($C_{i}=C$). \ Using MAPLE we can then derive expressions for the
intensity dependent dispersion, $R$, for Chain $\Lambda $ atoms of
varying number of states:
\begin{eqnarray*}
R^{(3)} &=&\frac{C^{2}}{\Omega ^{4}}, \\
R^{(5)} &=&\frac{C^{2}\left( \Omega ^{4}+2P^{2}\Omega ^{2}-2P^{4}\right) }{%
\left( \Omega ^{4}-P^{2}\Omega ^{2}+P^{4}\right) ^{2}}, \\
R^{(7)} &=&\frac{C^{2}\left( \Omega ^{8}+4P^{4}\Omega
^{4}-8P^{6}\Omega ^{2}+4P^{8}\right) }{\Omega ^{4}\left( \Omega
^{4}-2P^{2}\Omega
^{2}+2P^{4}\right) ^{2}}, \\
R^{(n)} &=&\frac{1}{P}\frac{\partial }{\partial \Delta _{p}}\left(
\sum_{i=1}^{n}\rho _{g_{i}e_{i}}\right) .
\end{eqnarray*}
We note that our results for $R^{(3)}$ are compatible with the
intensity dependent group velocities derived in Ref.
\cite{bib:Kuangpreprint}. \ The analytically determined
dispersions are plotted in Fig. 3 as a function of $P/\Gamma$ for
$C=0.25\Gamma $. \ In Fig. 4 we present a comparison between the
analytical results for the intensity dependent dispersion and that
obtained by solving the density matrix equation with decay given
in the Appendix. \ The good agreement shows that our analytic
approach is justified.

\begin{figure}[tb!]
\includegraphics[width=\columnwidth,clip]{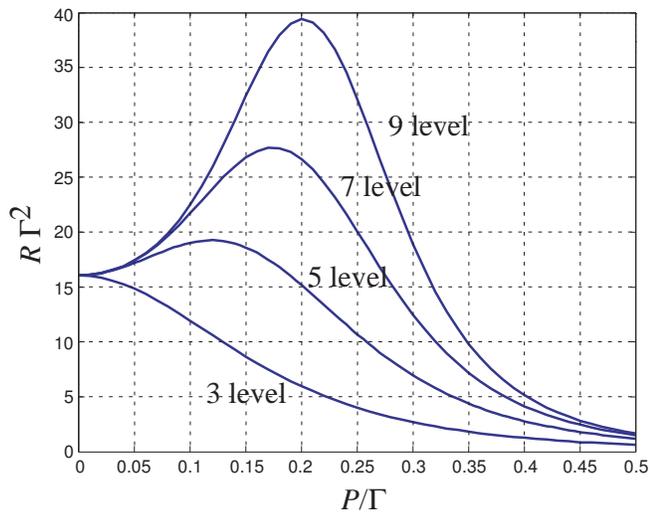}
\caption{\label{fig3} Graphs showing dispersion (times linewidth
squared), $R \Gamma^2 $ as a function of $P/\Gamma $ with $
C/\Gamma =0.25$ for Chain $\Lambda $ systems of 3,5,7 and 9
states.}
\end{figure}

\begin{figure}[tb!]
\includegraphics[width=\columnwidth,clip]{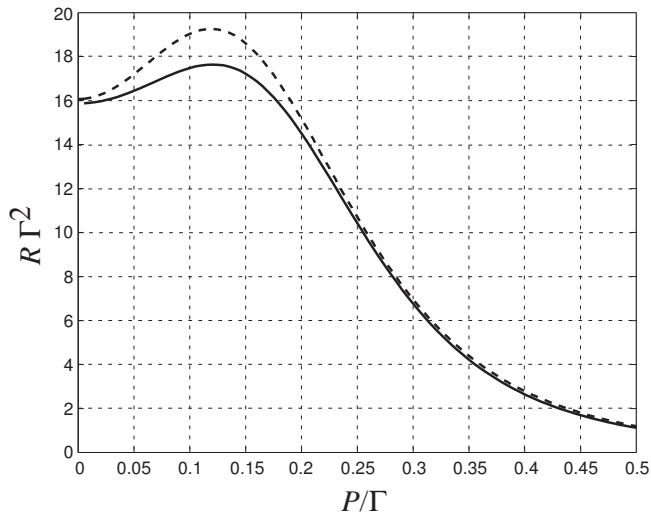}
\caption{\label{fig4} Graph showing the comparison between
intensity dependent dispersion calculated using the full master
equation (solid line) and the analytical approach using dressed
states (dashed line) in a five-state system.}
\end{figure}

If one simply performs a Taylor series expansion on our results
for the $R$'s (which are non-perturbative), it is easy to show
that the Chain $\Lambda$ systems exhibit nonlinearities to all
orders in $P/C$.  The linear dispersion has previously been
identified as being important in EIT systems (see for example
Harris {\it et al.} \cite{bib:HarrisPRA1992}), and it is clear
from Fig. 3 that the linear dispersion is identical for all Chain
$\Lambda$ atoms.  From this we may conclude that in the limit of
weak probe fields, the probe field cannot couple the levels in any
fashion other than the simple $\Lambda$ scheme. However as the
probe intensity is increased, higher-order process begin to turn
on.  In order to rigorously determine the order of the
nonlinearities present, one should construct an effective
Hamiltonian, following methods presented by, for example, Zubairy
{\it et al.} \cite{bib:Zubairy2002} or Klimov {\it et al.}
\cite{bib:KlimovQuantPh2002}.  This has not yet been performed for
the Chain $\Lambda$ system and it is hoped that such
investigations will shed more light on the nonlinear optical
properties of these systems.

An interesting feature to note in the dispersion calculations is
the Rabi frequency ratio which provides the maximum dispersion. \
In this simple case, the position of the maximum depends only upon
the ratio $P/C$ . \ If $\beta$ is the value of this ratio at the
maximum then for $P/C<\beta $ the dispersion will be monotonically
increasing with increasing $P$, and for $P/C>\beta $ it
monotonically decreases.  It is clear that reciprocal results will
be obtained for the corresponding group velocities. \ For $3$,
$5$, $7$ and $9$ state atoms, the values of $\beta $ are $0,$
$0.476,$ $0.698$ and $0.804$ (to 3 significant figures)
respectively.

One material property dependent on the dispersion is the group
velocity, which is
\[
v_{g}=\frac{c}{1+\omega _{p}\frac{\partial \Re (\eta) }{\partial
\Delta _{p}}},
\]
where $\eta=\sqrt{1+\chi}$ is the complex refractive index. It is
important to realize that in systems with nonlinear dispersions as
large as those for Chain $\Lambda$ systems, the group velocity may
be a poor parameter. This is because for realistic propagation
through an optically thick medium, the intensity dependence of the
medium will alter the shape of a simple gaussian pulse.  There are
however, other experiments sensitive to the group velocity which
may be considered, for example bichromatic excitation of the probe
beam to generate a beat note \cite{bib:MatskoReview2001} or the
use of a frequency modulated (rather than the more usual amplitude
modulated) probe signal.  In Fig. 5 we present intensity dependent
group velocities for $3$, $5$, $7$ and $9$ state Chain $\Lambda$
atoms corresponding to the dispersion calculations presented in
Fig. 3.

\begin{figure}[tb!]
\includegraphics[width=\columnwidth,clip]{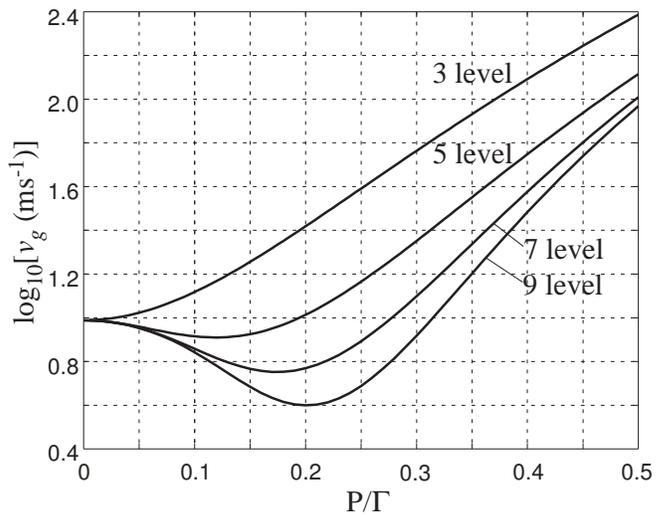}
\caption{\label{fig5} Graphs showing $\log_{10}$ of group
velocity, $v_g$ as a function of $P/\Gamma $ with $ C/\Gamma
=0.25$ for Chain $\Lambda $ systems of 3,5,7 and 9 states.}
\end{figure}

The analytical results given above cannot provide a complete
description of the physically realizable problem, because of the
different Clebsch-Gordan coupling between the states involved in
the transitions. \ If we use the scheme suggested in Fig. 2 for
the couplings and define our coupling
strengths relative to the coupling in the first $\Lambda $ system (i.e. $%
g_{1}-e_{1}-g_{2}$) then we may write down the expressions for the
other couplings in terms of these quantities \cite{bib:CGCalc}. \
These are summarized in table \ref{tab:Table3} for the 5 and 7
state Chain $\Lambda $ atoms, and it is easy to generalize to
higher orders.

\begin{table}
\begin{tabular}{||c||c|c|}
\hline\hline & 5-state & 7-state\\ \hline\hline $P_{2}$ & $\left(
\sqrt{3/2}\right) P_{1}$ & $\left( 2/\sqrt{2}\right) P_{1}$
\\
$C_{2}$ & $\left( \sqrt{2/3}\right) C_{1}$ & $\left(
6/\sqrt{30}\right)
C_{1} $ \\
$P_{3}$ & $-$ & $\left( \sqrt{15}/3\right) P_{1}$ \\
$C_{3}$ & $-$ & $\left( 3/\sqrt{15}\right) C_{1}$ \\
$\mu _{g_{2}e_{2}}$ & $\left( \sqrt{3/2}\right) \mu _{g_{1}e_{1}}$
& $\left(
2/\sqrt{2}\right) \mu _{g_{1}e_{1}}$ \\
$\mu _{g_{3}e_{3}}$ & $-$ & $\left( \sqrt{15}/3\right) \mu _{g_{1}e_{1}}$ \\
\hline
\end{tabular}
\caption{\label{tab:Table3} Relevant Rabi frequency ratios for 5
and 7 state Chain $\Lambda $ systems}
\end{table}

Results of numerical calculations of intensity-dependent
dispersions for 5- and 7-state Chain $\Lambda $ systems are
presented in Fig. 6. \ Comparing Figs. 3 and 6, shows that despite
the differences in the values obtained for the dispersions, there
is only minimal change to the overall {\em shape} of the intensity
dependent curves.

\begin{figure}[tb!]
\includegraphics[width=\columnwidth,clip]{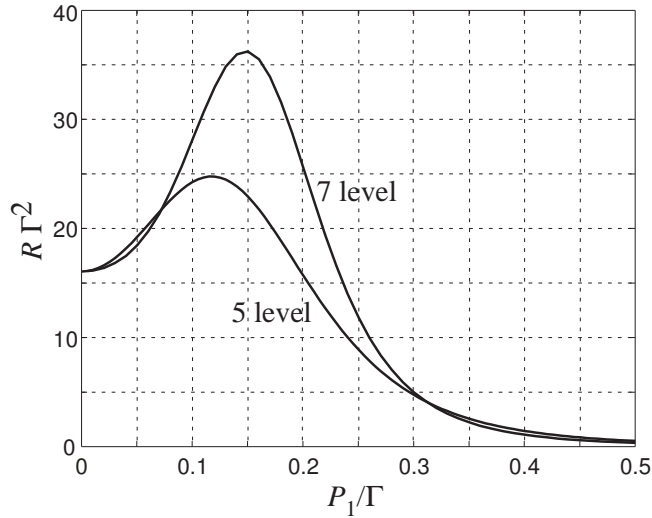}
\caption{\label{fig6} Graphs of intensity dependent dispersions as
a function of $P_1$, calculated using realistic ratios of coupling
strengths, i.e. with the relative $P_{i}$ and $ C_{i}$'s
calculated using the Clebsch-Gordan ratios in Table II.}
\end{figure}

We have shown that there exist interesting nonlinear properties
for Chain $ \Lambda $ atoms, and in particular we have focussed on
the intensity-dependent dispersion as a measure for these
nonlinear optical properties. \ The nonlinearity of these systems
increases as the number of atomic states increases, whilst the EIT
transparency is maintained. \ It therefore appears likely that
such multistate systems will be useful in the search for new
quantum non-linear optical materials. \ Our analysis has been
confined to the optically thin regime. \ Clearly, in a full study,
which would include intensity-dependent group velocities in such
highly nonlinear media, it is important to understand propagation
effects and especially the effect of such high nonlinearities on
pulse shape. \ Such analysis goes beyond our simple picture and
will be the focus of future work.

\section{Acknowledgements}
One of the authors (AG) would like to acknowledge useful
discussions with Dr A.B. Matsko (Jet Propulsion Laboratory) and
financial support from the EPSRC (UK).

\section{Appendix}

In this appendix we present the density matrix equations of motion
for the 5-state Chain $\Lambda $ atom ({\sf M} scheme). \ The
equations to be solved are:
\begin{eqnarray*}
\dot{\rho} &=&-\frac{i}{\hbar }\left[ {\cal H},\rho \right] +{\cal L}, \\
1 &=&\rho _{g_{1}g_{1}}+\rho _{g_{2}g_{2}}+\rho _{g_{3}g_{3}}+\rho
_{e_{1}e_{1}}+\rho _{e_{2}e_{2}}.
\end{eqnarray*}
For reasons of space we split the Hamiltonian superoperator into
smaller sub-blocks, thus
\[
{\cal H}=-\frac{i}{\hbar }\left(
\begin{array}{ccccc}
{\cal H}_{11} & {\cal H}_{12} & 0 & 0 & 0 \\
{\cal H}_{21} & {\cal H}_{22} & {\cal H}_{23} & 0 & 0 \\
0 & {\cal H}_{32} & {\cal H}_{33} & {\cal H}_{34} & 0 \\
0 & 0 & {\cal H}_{43} & {\cal H}_{44} & {\cal H}_{45} \\
0 & 0 & 0 & {\cal H}_{54} & {\cal H}_{55}
\end{array}
\right) ,
\]
where ${\cal H}_{2j,2j+1}=-C_{j}I\left( 5\right) $, ${\cal H}%
_{2j-1,2j}=-P_{j}$ and ${\cal H}_{ij}={\cal H}_{ji}$. $I\left(
5\right) $ is the $5\times 5$ identity matrix and
\begin{eqnarray*}
{\cal H}_{11} &=&\left(
\begin{array}{ccccc}
0 & P_{1} & 0 & 0 & 0 \\
P_{1} & \Delta _{p} & C_{1} & 0 & 0 \\
0 & C_{1} & \Delta_{cp} & P_{2} & 0 \\
0 & 0 & P_{2} & 2\Delta _{p}-\Delta _{c} & C_{2} \\
0 & 0 & 0 & C_{2} & 2 \Delta_{cp}
\end{array}
\right) , \\
{\cal H}_{22} &=&\left(
\begin{array}{ccccc}
-\Delta _{p} & P_{1} & 0 & 0 & 0 \\
P_{1} & 0 & C_{1} & 0 & 0 \\
0 & C_{1} & -\Delta _{c} & P_{2} & 0 \\
0 & 0 & P_{2} & \Delta_{cp} & C_{2} \\
0 & 0 & 0 & C_{2} & \Delta _{p}-2\Delta _{c}
\end{array}
\right) , \\
{\cal H}_{33} &=&\left(
\begin{array}{ccccc}
-\Delta _{p} & P_{1} & 0 & 0 & 0 \\
P_{1} & 0 & C_{1} & 0 & 0 \\
0 & C_{1} & -\Delta _{c} & P_{2} & 0 \\
0 & 0 & P_{2} & \Delta_{cp} & C_{2} \\
0 & 0 & 0 & C_{2} & \Delta _{p}-2\Delta _{c}
\end{array}
\right) , \\
{\cal H}_{44} &=&\left(
\begin{array}{ccccc}
-2\Delta _{p}+\Delta _{c} & P_{1} & 0 & 0 & 0 \\
P_{1} & -\Delta _{cp} & C_{1} & 0 & 0 \\
0 & C_{1} & -\Delta _{p} & P_{2} & 0 \\
0 & 0 & P_{2} & 0 & C_{2} \\
0 & 0 & 0 & C_{2} & -\Delta _{c}
\end{array}
\right) , \\
{\cal H}_{55} &=&\left(
\begin{array}{ccccc}
-2 \Delta_{cp}  & P_{1} & 0 & 0 & 0 \\
P_{1} & -\Delta _{p}+2\Delta _{c} & C_{1} & 0 & 0 \\
0 & C_{1} & -\Delta_{cp} & P_{2} & 0 \\
0 & 0 & P_{2} & \Delta _{c} & C_{2} \\
0 & 0 & 0 & C_{2} & 0
\end{array}
\right) .
\end{eqnarray*}
where $ \Delta_{cp} = \Delta_P - \Delta_C$.  The loss operator is
\[
{\cal L}=\left(
\begin{array}{ccccc}
{\cal L}_{11} & {\cal L}_{12} & 0 & 0 & 0 \\
0 & {\cal L}_{22} & 0 & 0 & 0 \\
0 & {\cal L}_{32} & {\cal L}_{33} & {\cal L}_{34} & 0 \\
0 & 0 & 0 & {\cal L}_{44} & 0 \\
0 & 0 & 0 & {\cal L}_{54} & {\cal L}_{55}
\end{array}
\right) ,
\]
where the off-diagonal blocks ${\cal L}_{ij}$ $\left( i\neq
j\right) $ have
all elements zero except the $\left( ij\right) ^{\text{th}}$ which is $%
\Gamma /2$. \ The ${\cal L}_{ii}$'s are all diagonal matrices,
with non-zero elements
\begin{eqnarray*}
{\cal L}_{11} &=&\left( 0,-\Gamma /2,-\Gamma _{2},-\Gamma
_{3},-\Gamma
_{4}\right) , \\
{\cal L}_{22} &=&\left( -\Gamma /2,-\Gamma ,-\Gamma /2,-\Gamma
,-\Gamma
_{3}\right) , \\
{\cal L}_{33} &=&\left( \Gamma _{2},-\Gamma /2,0,-\Gamma
/2,-\Gamma
_{2}\right) , \\
{\cal L}_{44} &=&\left( -\Gamma _{3},-\Gamma ,-\Gamma /2,-\Gamma
,-\Gamma
/2\right) , \\
{\cal L}_{55} &=&\left( -\Gamma _{4},-\Gamma _{3},-\Gamma
_{2},-\Gamma /2,0\right) ,
\end{eqnarray*}
where $\Gamma _{n}$ is the $n$ photon dephasing and $\Gamma
=\Gamma _{e_{1}}=\Gamma _{e_{2}}$ is the total decay rate from
either excited state.

A three-dimensional plot showing $-\Im(\chi)$ as a function of
probe detuning, $\Delta_p / \Gamma$ and probe Rabi frequency, $P/
\Gamma$ for $P_1 = P_2 = P$, $C_1 = C_2 = C = \Gamma / 4$, $\Gamma
= 5.6 {\rm MHz}$ and other parameters as above, for the five-state
Chain $\Lambda$ system is presented in Fig. 7.

\begin{figure}[tb!]
\includegraphics[width=\columnwidth,clip]{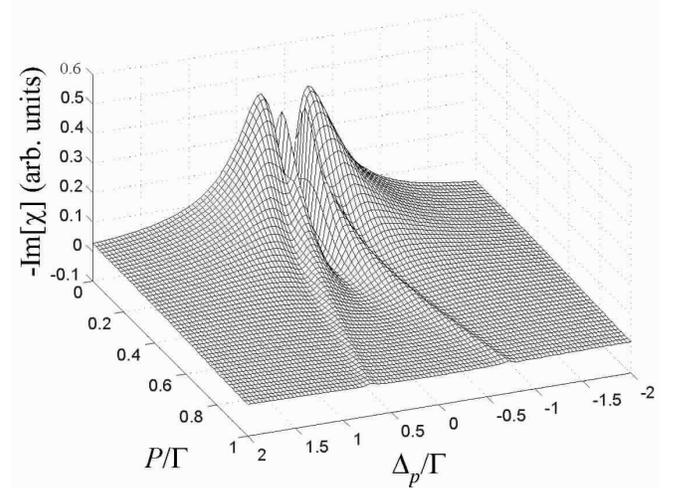}
\caption{\label{fig7} Three-dimensional plot showing $-\Im[\chi]$
(proportional to the probe absorption) as a function of $\Delta_p
/ \Gamma$ and $P/ \Gamma$ for $P_1 = P_2 = P$, $C_1 = C_2 = C =
\Gamma / 4$, $\Gamma = 5.6 {\rm MHz}$ for the five-state Chain
$\Lambda$ system.}
\end{figure}

\end{document}